\documentclass[showpacs,amssymb,eqsecnum,twocolumn,aps]{revtex4}
\usepackage{amsmath}
\usepackage{graphicx}
\usepackage{epsf}
\usepackage{graphics}
\usepackage[latin1]{inputenc}
\newcommand{\be}{\begin{equation}}
\newcommand{\ee}{\end{equation}}
\newcommand{\bea}{\begin{eqnarray}}
\newcommand{\eea}{\end{eqnarray}}

\newcommand{\teb}[1]{\vec{#1}}

\newcommand{\rmd}{{\rm d}}

\DeclareRobustCommand{\teb}[1]{
 \begingroup
 \mathchoice
   {\hbox{{$\displaystyle\oalign{$#1$\crcr\hidewidth\vbox
            to.2ex{\hbox{\LARGE{\char126}}\vss}\hidewidth}  $}}}%
   {\hbox{{$\textstyle\oalign{$#1$\crcr\hidewidth\vbox
            to.2ex{\hbox{\LARGE{\char126}}\vss}\hidewidth}  $}}}%
   {\hbox{{$\scriptstyle\oalign{$\scriptstyle #1$\crcr\hidewidth\vbox
            to.2ex{\hbox{\char126}\vss}\hidewidth}  $}}}%
   {\hbox{{$\scriptscriptstyle
            \oalign{$\scriptscriptstyle #1$\crcr\hidewidth\vbox
            to.2ex{\hbox{\char126}\vss}\hidewidth}  $}}}
 \endgroup}
\newcommand{\unf}{\teb{F}}
\newcommand{\ung}{\teb{G}}
\newcommand{\una}{\teb{A}}
\newcommand{\unm}{\teb{M}}
\newcommand{\unn}{\teb{N}}

\begin{document}

\title{Resummation of infrared divergences in the free-energy
  of spin-two fields}

\author{F. T. Brandt$^{a}$, J. Frenkel$^{a}$, D. G. C. McKeon$^{b}$
and J. B. Siqueira$^{a}$} 
\affiliation{$^a$ Instituto de F\'{\i}sica,
Universidade de S\~ao Paulo,
S\~ao Paulo, SP 05315-970, Brazil}
\affiliation{$^b$ Department of Applied Mathematics, University of
Western Ontario, London, ON  N6A 5B7, Canada}

\begin{abstract}
We derive a closed form expression for the sum of all the infrared 
divergent contributions to the free-energy of a gas of
gravitons. An important ingredient of our calculation is the use of a
gauge fixing procedure such that the graviton propagator becomes both
traceless and transverse. This has been shown to be possible,
in a previous work, using a general gauge fixing procedure,
in the context of the lowest order expansion of the Einstein-Hilbert action,
describing non-interacting spin two fields.
In order to encompass the problems involving thermal loops, such as
the resummation of the free-energy, in the present work, we
have extended this procedure to the situations when the interactions
are taken into account. 
\end{abstract}

\maketitle

\section{Introduction}
The usual perturbative expansion of the free-energy of a gas of massless bosons
at temperature $T$
contains infrared divergences when three or more loops are taken into account. 
When all the infrared divergent diagrams are summed, the resulting
expression is finite but
non-analytic in the coupling constant. This is well known
in the context of scalar or spin-one 
gauge fields \cite{kapusta:book89,lebellac:book96,das:book97}.
The same situation is expected for spin-two gauge fields when we 
apply the finite temperature field theory techniques to the graviton
gauge field described by the weak field expansion of the Einstein-Hilbert action
\cite{tHooft:2002xp}. One of the main purposes of the present
paper is to obtain the explicit full result for the summation of these so
called {\it ring diagrams} in the case of gravity.

\begin{figure}
\includegraphics[scale=0.6]{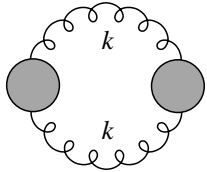}
\caption{The lowest order infrared divergent contribution to the
  free-energy. The curly line represents the graviton propagator 
$D_{\mu\nu,\,\alpha\beta}$ and the blob represents the
graviton self-energy.}\label{fig0}
\end{figure}

In figure \ref{fig0} we show a
typical infrared divergent ring diagram containing two insertions of the graviton
self-energy.
It illustrates the two important ingredients in the analysis of any of
the higher order ring contributions.
First we need the full tensor structure of the 
dominant high temperature contribution to the 
graviton self-energy (the blob in figure \ref{fig0})
in the static limit. This is a well known quantity
which has been studied and shown to be gauge independent
\cite{Rebhan:1990yr,Brandt:1998hd}. 
Secondly, we need the three-level graviton propagator (the curly line
in figure \ref{fig0}) connecting 
the two self-energies in the ring diagram.
Once we know these quantities any ring diagram can be obtained by
multiple insertions of the self-energy in a closed loop of gravitons,
in a rather straightforward manner. At this point the tensor
properties of the free-propagator are essential in order to obtain a
closed form and simple result for the resummed free-energy. As we will
see, if the propagator satisfies the following traceless-transverse (TT) conditions
\begin{subequations}\label{ttcond}
\begin{eqnarray}\label{e6}
\eta^{\mu\nu} D^{\rm TT}_{\mu\nu,\,\lambda\sigma}(k) &=&0 \\
\label{e7}
k^{\mu} D^{\rm TT}_{\mu\nu,\,\lambda\sigma}(k) &=&0 ,
\end{eqnarray}
\end{subequations}
then the sum of the ring diagrams acquire a simple form in terms of
the three TT projections of the static high-temperature limit of the
graviton self-energy.

Taking into account the gauge independence of the leading contributions
of the ring diagrams, one can choose the most convenient gauge for
the graviton propagator. It has been shown in
Ref. \cite{Brandt:2007td} that it is possible to choose a gauge such
that the graviton propagator becomes TT.
In principle we could just assume that the known gauge
invariant result for the graviton self-energy could be used safely in
the ring diagrams. However, since the derivation of the TT propagator is
only possible if we consider some modifications in the
usual Faddeev-Popov procedure, which necessarily leads to new ghosts and
interactions, we have also performed the explicit
calculation of the leading static graviton self-energy and verified
that the result is indeed the correct gauge invariant one. Considering
that this calculation involves some rather nontrivial cancellations of diagrams,
it also constitutes an important test
of the gauge fixing procedure introduced in \cite{Brandt:2007td}.
%and  applied in the rather non-trivial calculations of the present work.

This paper is organized as follows.
In section II we will review the generalization of the Faddeev-Popov
formalism which leads to a traceless-transverse graviton
propagator. We extend the procedure presented in reference 
\cite{Brandt:2007td} by taking into consideration 
interactions of gravitons and ghosts more explicitly.
In section III we present the calculation of the
high temperature static limit of the graviton self-energy. 
Having verified that our gauge fixing procedure yields the correct
gauge independent result for the self-energy, we proceed, in section IV, to the 
calculation of the sum of all the infrared divergent contributions to
the free-energy. Finally, in section V we discuss the main results.

\section{General gauge fixing and Feynman rules}
In this section we will follow the basic idea of
section II of Ref. \cite{Brandt:2007td} in order to derive the TT
propagator. In addition to that derivation, we will also obtain 
the interaction vertices involving three types of ghosts and gravitons
(see also the reference \cite{brandt:087702} for an analogous derivation in the
context of spin-one fields were it has been shown that a global gauge
invariance, analogous to the BRST invariance is present in the
effective action). 
This involves a generalization of the
well known Faddeev-Popov procedure \cite{Faddeev:1967fc,Hooft:1971fh}
to the cases when the gauge fixing condition is non-quadratic.

In order to illustrate the main features of the generalization of the 
Faddeev-Popov procedure, it is convenient to
consider the following integral over the components of a
$n$-dimensional vector $\vec h$
\be\label{eq5}
Z = \int {\rm d}\vec h \exp{S(\vec h)},
%=\frac{\pi^{\frac{n}{2}}}{(\det\unm)^{\frac{1}{2}}}.
\ee
where 
\be\label{eq5a}
S(\vec h) = -\vec h^T\unm\vec h + S_i(\vec h).
\ee
Later we will associate $\vec h$ with the graviton field $h_{\mu\nu}(x)$;
the first and second terms Eq. \eqref{eq5a} will be identified respectively with 
the quadratic and interaction terms 
which arise from the weak field expansion of the Einstein-Hilbert
action. 

Let us now consider the interesting case when $S(\vec h)$ is invariant under an
infinitesimal transformation of the form
\be
\vec h \rightarrow \vec h + \una(\vec h) \vec\theta,
\ee
where the operator $\una(\vec h)$ is of first order 
in the derivative operator $\partial$ as well as in $\vec h$ .
%and terms containing ${\cal O}(\vec\theta)^2)$ can be neglected. The quantity 
%$\vec \theta$ is the generator of the gauge transformation.
This symmetry 
makes the integration in Eq. \eqref{eq5} undefined so that we have to
employ the Faddeev-Popov procedure which leads to the introduction of
a ``gauge fixing term'', yielding a
quadratic term of the form $\vec h \teb{\bar{M}}\vec h$,
such that $\teb{D}^{(0)}={\teb{\bar{M}}}^{-1}$ is well defined.
In the context of gauge field theories, $\teb{D}^{(0)}$ is the
free propagator which will be dependent on the specific choice of
gauge fixing.  In the case of gravity, some rather
general gauge fixing conditions have been investigated previously
\cite{Nishino:1977pw}. However, in the Ref. \cite{Brandt:2007td} 
it has been shown that it is not possible
to obtain a graviton propagator satisfying the TT conditions given in 
\eqref{ttcond}, using the standard gauge fixing procedures.
In what follows we will present the main steps which are involved
in the generalized gauge fixing and them apply the results to the case
of the Einstein-Hilbert action.

First, we introduce the following factors of ``$1$'' in the integrand of Eq. \eqref{eq5}
% (4.69) tese joão
\begin{subequations}\label{eq6}
\bea\label{eq6a}
1=\int{\rm  d}\vec\theta_1\delta(\unf(\vec h+\alpha\una\vec\theta_1))
\det(\alpha\unf\una)
\\ \label{eq6b}
1=\int{\rm  d}\vec\theta_2\delta(\ung(\vec h+\alpha\una\vec\theta_2))
\det(\alpha\ung\una)
\eea
and
\bea\label{eq8}
1=\left(\alpha\pi\right)^{-n}\int{\rm d}\vec p\; {\rm d}\vec
q\exp{\left(-\frac{1}{\alpha}\vec p^T\unn\vec q\right)}\det{\unn}.
\eea
\end{subequations}
where the $\unn$ is a ``Nielsen-Kallosh'' 
factor  \cite{DeWitt:1967yk,Nielsen:1978mp,Kallosh:1978de},
$\unf$ and $\ung$ are two independent operators, which we will assume
that are of first order in $\partial$, and $\delta$ is the Dirac delta function.
The use of two operators makes the gauge fixing prescription 
more general; 
%so that we will be able to obtain a TT propagator. 
the usual Faddeev-Popov procedure would introduce only two factors of
``$1$'', which corresponds to make the special identification $\unf=\ung$.

The next step consists in using the infinitesimal gauge transformation
\be\label{eq9}
\vec h \rightarrow \vec h -\alpha \una \vec \theta_{1}.
\ee
After integrating out the  $\vec p$ and $\vec q$ variables
and setting, for simplicity, $\unn$ equal to the identity operator, we obtain
%\begin{widetext}
\bea\label{eq10}
Z&=&\left({\alpha}\right)^{2n}
\int{\rm  d}\vec\theta_1\int{\rm  d}\vec\theta\int{\rm  d}\vec h
\det(\unf\una)
\det(\ung\una)
\nonumber \\
&\times& 
\exp\left[-\vec
    h^T\left(\unm+\frac{1}{\alpha}\unf^T \ung\right)\vec h 
\right. \nonumber \\  & & 
\left. \qquad \qquad \qquad 
+{S}_i(\vec h)
-\vec h^{T}\unf^T \ung\una\vec\theta\right]
\eea 
%\end{widetext}
where $\vec\theta=\vec\theta_{2} - \vec\theta_{1}$. 

If we want to avoid the use of mixed propagators which would be
generated by the last term in \eqref{eq10}, then we also have to
perform the shift 
\be\label{e65}
\vec h \rightarrow 
\vec h +\frac{1}{2}
\left(\teb{M} + \frac{1}{\alpha}
     \teb{F}^{T}\teb{G}\right)^{-1}
\left(\teb{F}^{T}\teb{G}\teb{A}\right)
\vec\theta ,
\ee
so that \eqref{eq10} yields
\begin{eqnarray}\label{e66}
Z & = & 
\int{\rm d}\vec\theta \int{\rm d}\vec h 
\det(\teb{F}\teb{A})
\det(\teb{G}\teb{A}) 
J
\nonumber \\
&\times & 
 \exp\left\{-\vec h^T \left(\teb{M} + \frac{1}{\alpha}
\teb{F}^{T}\teb{G}\right)\vec h
\right. \nonumber \\ &+& 
{S}_i\left(
\vec h +\frac{1}{2}
\left(\teb{M} + \frac{1}{\alpha}
     \teb{F}^{T}\teb{G}\right)^{-1}
\left(\teb{F}^{T}\teb{G}\teb{A}\right)
\vec\theta 
\right) \left.
\right. \nonumber \\ &-& \left.
\frac{1}{4}
\vec\theta^T
\left({\teb{A}}^{T}\teb{G}^{T}\teb{F}\right)
\left(\teb{M} + \frac{1}{\alpha}
     \teb{F}^{T}\teb{G}\right)^{-1}
\right. \nonumber \\ &\times& \left.
\left(\teb{F}^{T}\teb{G}\teb{A}\right)
\vec\theta
\right\},
\nonumber \\
\end{eqnarray}
where we have dropped the infinite normalization factors as well as the
integration over the gauge orbit $\int {\rm d} \vec\theta_1$. Since
the transformation \eqref{e65} is not a gauge transformation, its
Jacobian may not be equal to one. For this reason, we have also
introduced the Jacobian factor $J$ in the integrand of \eqref{e66}.

The determinants in Eq. \eqref{e66} can be exponentiated using the
standard Berezin integral 
\be\label{eq11}
\det\teb{B} = \int{\rm d}\vec c\;{\rm d}\vec{\bar{c}}
\;\exp(-\vec{\bar{c}}^{\,T}\teb{B}\vec c)
\ee 
where $\vec c$, $\vec{\bar{c}}$ are Grassmann vectors; the first two
determinants in Eq. \eqref{eq10} lead to ``Faddeev-Popov" like ghosts
and the  field $\vec\theta$ is a ``Bosonic'' ghost. There would be
also an extra ghost field associated with the determinant
$J$. However, one can argue that in the applications to be considered in the present
work, these extra ghost fields will not contribute.
The final expression for $Z$ can be written as
\be
Z = \int \rmd \vec\theta\,
{\rm d}\vec c\;{\rm d}\vec{\bar{c}}\;
{\rm d}\vec d\;{\rm d}\vec{\bar{d}} \;\rmd\vec h
\,J\, \exp{\left[S_{eff} %(\vec\theta,\vec c, \vec d, \vec h)
\right]},
\ee
where 
\begin{widetext}
\begin{eqnarray}\label{seff1}
{S_{eff}}
%(\vec h, \vec c , \vec d , \vec\theta)} 
&=&
-\vec h^T \left(\teb{M} + \frac{1}{\alpha}
\teb{F}^{T}\teb{G}\right)\vec h
+{S}_i\left(
\vec h +\frac{1}{2}
\left(\teb{M} + \frac{1}{\alpha}
     \teb{F}^{T}\teb{G}\right)^{-1}
\left(\teb{F}^{T}\teb{G}\teb{A}\right)
\vec\theta 
\right) 
\nonumber \\ &-& 
\frac{1}{4}
\vec\theta^T
\left({\teb{A}}^{T}\teb{G}^{T}\teb{F}\right)
\left(\teb{M} + \frac{1}{\alpha}
     \teb{F}^{T}\teb{G}\right)^{-1}
\left(\teb{F}^{T}\teb{G}\teb{A}\right)
\vec\theta
-\vec{\bar{c}}^{\,T}{\unf\una}\vec c-
\vec{\bar{d}}^{\,T} {\ung\una}\vec d .
\end{eqnarray}
\end{widetext}

Let us now consider the specific example when the general ``action'' in
\eqref{seff1} describes the behaviour of spin-two fields.
In this case, the classical dynamics is obtained from the
Einstein-Hilbert action
\begin{equation}\label{seh}
S_{EH}=\frac{2}{\kappa^2}\int \rmd^{d}x \sqrt{-g} R, 
\end{equation}
where $\kappa^2={32 \pi G }$ ($G$ has mass dimension $2-d$)
$R$ is the Ricci scalar, $g=\det g_{\mu\nu}$ and
\be
g_{\mu\nu}=\eta_{\mu\nu}+\kappa h_{\mu\nu}
\ee
is the definition of the metric in terms of the graviton field
$h_{\mu\nu}$, which is to be associated with $\vec h$.
It is straightforward to obtain the corresponding expression for 
the quadratic operator $\teb{M}$ as well as the first
interaction terms (there will be an infinity number of graviton
self-interactions) when \eqref{seh} is expanded in powers of the $\kappa$.
From the lowest order quadratic contribution one obtains the following
tensor components of $\teb{M}$
\begin{eqnarray}\label{m11}
(\teb{M})^{\mu \nu, \alpha \beta}&=&-\frac{1}{2}\left[\frac{1}{2}(\eta^{\mu
    \alpha}\eta^{\nu \beta}+\eta^{\mu \beta}\eta^{\nu
    \alpha})-\eta^{\mu \nu}\eta^{\alpha \beta}\right] \partial^2
\nonumber  \\  
&&\!\!\!\!\!\!\!\!\!\!\!\!\!\!\!\!\!\!\!\!\!\!\!\!+
\frac{1}{4}(\eta^{\mu
  \alpha}\partial^{\nu}\partial^{\beta} 
+ \eta^{\mu \beta}\partial^{\nu}\partial^{\alpha} 
+ \eta^{\nu \alpha}\partial^{\mu}\partial^{\beta} 
+ \eta^{\nu \beta}\partial^{\mu}\partial^{\alpha})
\nonumber \\ 
&& \!\!\!\!\!\!\!\!\!\!\!\!\!\!\!\!\!\!\!\!\!\!\!\!-
\frac{1}{2}(\eta^{\mu \nu}\partial^\alpha \partial^\beta+
\eta^{\alpha \beta} \partial^\mu \partial^\nu )
\end{eqnarray}

The Einstein-Hilbert action \eqref{seh} is invariant
under the space-time dependent coordinate transformation 
$x^\mu\rightarrow x^\mu - \theta^\mu(x)$, where
$\theta^\mu(x)$ are the generators of the
transformation. This induces the following gauge transformation of the graviton field
\be\label{transf}
\kappa h_{\mu \nu}
\rightarrow \kappa h_{\mu \nu}+(\una)_{\mu \nu , \lambda}(h) \theta^\lambda,
\ee
where
\begin{eqnarray}\label{a11}
(\una)_{\mu \nu , \lambda}&=&\eta_{\mu \lambda}\partial_\nu +\eta_{\nu \lambda}\partial_\mu
\nonumber \\ &+&
\kappa\left(h_{\mu \lambda}\partial_\nu +h_{\nu \lambda}\partial_\mu + (\partial_\lambda h_{\mu \nu})\right)
\end{eqnarray}
can now be identified with the
tensor components of $\una$ in \eqref{seff1}.

Let us now introduce the tensor components of the gauge fixing conditions 
$\unf$ and $\ung$. Following the analysis performed in 
Ref. \cite{Brandt:2007td} we choose
\begin{subequations}\label{f11}
\begin{eqnarray}
(\unf)^{\lambda , \mu \nu}&=& g_1 \eta^{\mu \nu} \partial^\lambda
+\eta^{\mu \lambda} \partial^{\nu} 
\\
(\ung)^{\lambda , \mu \nu}&=& g_2 \eta^{\mu \nu} \partial^\lambda +\eta^{\mu \lambda} \partial^{\nu},
\end{eqnarray}
\end{subequations}
where $g_1$ and $g_2$ are two independent gauge parameters.
It has been shown that this choice is sufficiently general 
in order to make the propagator TT when
the limit $\alpha\rightarrow 0$ is taken. In addition, the conditions
defined by \eqref{f11} also 
interpolate continuously between other usual gauges such
as the de Donder gauge in which case $g_1=g_2=-1/2$ and $\alpha=1$.

From the explicit expressions for $\teb{M}$, $\una$, $\unf$ and $\ung$
given respectively by Eqs. \eqref{m11}, \eqref{a11} and \eqref{f11},
together with Eq. \eqref{seff1} and the relevant expressions for
${S}_{int}$, the momentum space Feynman rules can now be obtained.
Notice also that the multiplication rules for the operators 
$\unf$, $\ung$ and $\una$ are such that
\be
(\unf^T \ung)^{\mu\nu,\,\alpha\beta} = 
F_{\lambda ,}^{\;\;\mu \nu}\, G^{\lambda , \alpha \beta},
\ee
\be
(\ung \una)^{\delta}_{\;\lambda} = G^{\delta,\mu\nu} A_{\mu\nu ,\lambda} ,
\ee
\be
(\unf^T \ung \una)^{\mu\nu}_{\;\;\; \lambda}=
F_{\delta,}^{\;\;\mu \nu}\, G^{\delta , \alpha \beta}
A_{\alpha\beta,\,\lambda} 
\ee
with analogous relations for the other terms in
Eq. \eqref{seff1}. Also, the individual terms in the action are such that,
for instance,
\be
{\vec h}^T \teb M \vec h = \int \rmd^d x\, h_{\mu\nu} 
(\teb{M})^{\mu\nu\,\alpha\beta} h_{\alpha\beta} 
\ee
and
\be
\vec {\bar c} \,(\unf \una)\,\vec c  = \int \rmd^d x\, 
{\bar c}^\lambda (\unf \una)^\delta_{\;\lambda} c_\delta.
\ee

The momentum space Feynman rules can now be derived in the usual
fashion using the above identifications for each term
in the action \eqref{seff1}. (This tedious but straightforward task 
has been done with the help of the computer algebra program HIP \cite{hsieh1:1992ti}.)
The expressions for the graviton and the $\theta$ ghosts propagators are explicitly
given in Ref. \cite{Brandt:2007td}. For completeness (and also to
define the present normalizations and notations) let us briefly rederive
those expressions. Here we will also give the explicit expressions for
the $c$ and $d$ ghost propagators.

The free graviton propagator in momentum space,
${D}^{h}_{\mu \nu ; \alpha \beta}(k)$, can be readily obtained
from the the first term in Eq. \eqref{seff1} as the solution of
\begin{equation}\label{defprop}
{D}^{h}_{\mu \nu ; \alpha \beta}
{\left(\teb{M}+\frac{1}{\alpha}{\unf^T \ung}\right)}^{\alpha
  \beta ; \lambda \gamma}_{\mbox {\scriptsize symm.}} = I_{\mu \nu}^{\lambda \gamma},
\end{equation} 
where the subscript ``symm'' indicates that we have taken into account
the Bosonic symmetry of the graviton field $h_{\mu\nu}$. Also, 
it is implicit that we
have already made the Fourier transformations to the momentum space.
From the symmetry properties under interchange of tensor indices,
it follows that we can parametrize the solution of Eq. \eqref{defprop} 
in terms of five quantities ${\bf C}_i$, $i=1,\cdots 5$ as 
\be\label{grav_prop_gen}
{D}^{h}_{\mu \nu ; \alpha \beta}(k) =  \frac{1}{2 k^2}
\sum_{i=1}^5 {\bf C}^i T^i_{\mu\nu,\,\alpha\beta}(k),
\ee
where
\begin{eqnarray}\label{base1}
T^1_{\mu \nu \alpha \beta}(k)&=& \eta_{\mu \alpha} \eta_{\nu \beta} + \eta_{\mu \alpha} \eta_{\nu \beta} \nonumber\\
T^2_{\mu \nu \alpha \beta}(k)&= &\eta_{\mu \nu} \eta_{\alpha \beta}\nonumber\\
T^3_{\mu \nu \alpha \beta}(k)&=& \frac{1}{k^2}
(\eta_{\mu \alpha} k_{\nu} k_{\beta}+\eta_{\mu \beta} k_{\nu}k_{\alpha})+\mu\leftrightarrow\nu
\nonumber \\
T^4_{\mu \nu \alpha \beta}(k)&=& \frac{1}{k^2}(\eta_{\mu \nu}
k_{\alpha} k_{\beta}+\eta_{\alpha \beta} k_{\mu} k_{\nu}) \\
T^5_{\mu \nu \alpha \beta}(k)&=& \frac{1}{k^4}k_{\mu}
k_{\nu}k_{\alpha} k_{\beta}. \nonumber
\end{eqnarray}
In terms of this tensor basis, the solutions for ${\bf C}^i$
are given by Eqs. (48) of Ref. \cite{Brandt:2007td}.
They depend on $g_1$, $g_2$, $\alpha$ and the space-time dimension
$d$, in such a way that the TT property \eqref{ttcond} is fulfilled
when the limit $\alpha\rightarrow 0$ is taken. In this case, the
propagator can be expressed as
\begin{eqnarray}\label{propTT}
{D}^{TT}_{\mu \nu ; \alpha \beta} &=& \frac{1}{k^2}\left(
 \frac{1}{2} P_{\mu \alpha} P_{\nu \beta} 
+\frac{1}{2} P_{\nu \alpha} P_{\mu \beta} \right.
\nonumber \\
&-&\left.\frac{1}{d-1} P_{\mu \nu} P_{\alpha \beta}\right)
\end{eqnarray}
where
\begin{equation}
P_{\mu \nu}=\eta_{\mu \nu } -\frac{k_\mu k_\nu}{k^2}
\end{equation}
(notice that the transversality $k^{\mu} P_{\mu \nu} =0$,
idempotency $P_{\mu \alpha} P^{\alpha}_{\nu} = P_{\mu \nu}$ 
and the trace $\eta_{\mu \nu} P^{\mu \nu} = d-1$ 
guarantee the TT condition \eqref{ttcond}).

Let us now consider the quadratic ghost sectors of the action.
From the last three terms in Eq. \eqref{seff1} one can obtain the quadratic
and interaction terms for the three ghost fields. The quadratic terms can be
readily obtained considering the contribution of the first two terms in
\eqref{a11}. From these quantities the ghost
propagators associated to the fields $c^\mu$ and $d^\mu$ are given respectively by
\begin{subequations}
\be
D^c_{\mu \nu}(k)= \frac{ (2 g_1+1) (\frac{k_\mu k_\nu}{k^2} -\eta_{\mu
    \nu}) - \eta_{\mu \nu}}{{2(g_1+1)} k^2}
\ee
and
\be
D^d_{\mu \nu}(k)= \frac{ (2 g_2+1) (\frac{k_\mu k_\nu}{k^2} -\eta_{\mu
    \nu}) - \eta_{\mu \nu}}{{2(g_2+1)} k^2}
\ee
\end{subequations}
Similarly the $\theta$ sector of \eqref{seff1} yields the following
expression for the propagator of the $\theta$ ghost
\begin{eqnarray}\label{theta_prop}
{D^\theta}_{\mu \nu}&=& \frac{2}{\alpha k^4} \left\{
  \eta_{\mu\nu}-
\left[1-\frac{1}{4(g_1+1)(g_2+1)}
  \right. \right.  
\nonumber \\ &\;& 
\left. \left. - \frac{1(d-1)}{8\alpha(d-2)}\left(\frac{1}{g_1+1}-\frac{1}{g_2+1}\right)^2\right]\frac{k_{\mu}k_{\nu}}{k^2}\right\} 
\nonumber \\ & &
\end{eqnarray}

\begin{figure}
\[
\begin{array}{cc}
\includegraphics[scale=0.5]{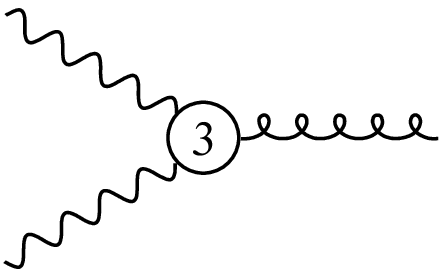} & \includegraphics[scale=0.5]{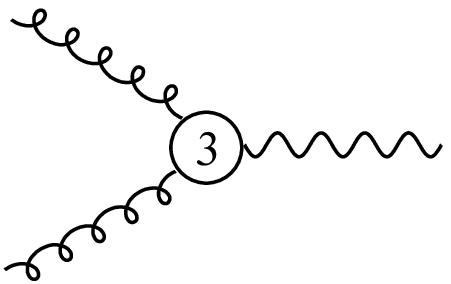}
\\ & \\ (a) & (b) \\ & \\
\includegraphics[scale=0.5]{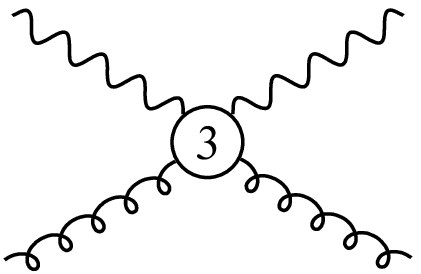} & \includegraphics[scale=0.5]{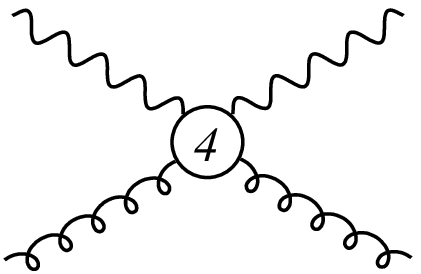}
\\ & \\ (c) & (d) \\ & 
\end{array}
\]
\caption{Interactions between the graviton and the $\theta$
ghost arising from the second term in Eq. \eqref{seff1}. 
The graviton and the ghost fields are respectively depicted
by curly and wavy lines. Graphs (a), (b) and (c) arise from the
shifted cubic term and graph (d) from the shifted quartic term.}\label{f1}
\end{figure}

\begin{figure}
\[
\begin{array}{cc}
\includegraphics[scale=0.5]{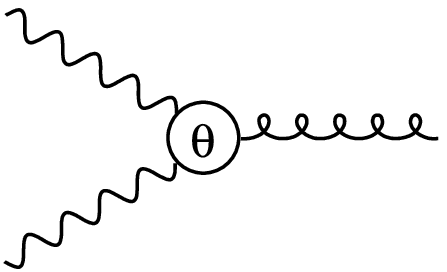} & \includegraphics[scale=0.5]{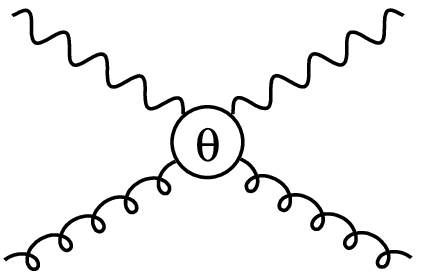}
\\ & \\ (a) & (b) \\ &
\end{array}
\]
\caption{Interactions between the graviton and the $\theta$
ghost arising from the third term in Eq. \eqref{seff1}.}\label{f2}
\end{figure}

\begin{figure}
\[
\begin{array}{cc}
\includegraphics[scale=0.5]{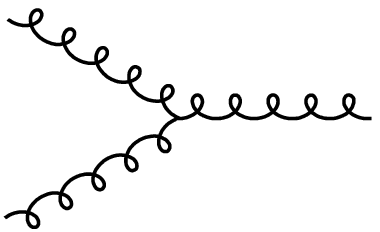} & \includegraphics[scale=0.5]{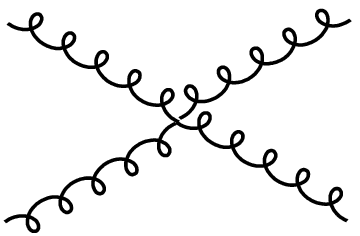}
\\ & \\ (a) & (b) \\ & \\
\includegraphics[scale=0.5]{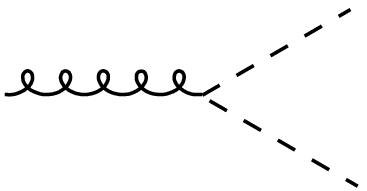} & \includegraphics[scale=0.5]{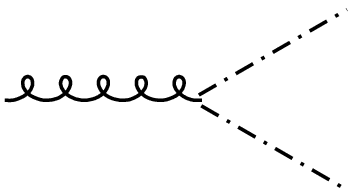}
\\ & \\ (c) & (d) \\ &
\end{array}
\]
\caption{Figures (a) and (b) represent the graviton 
self-interactions from the Einstein-Hilbert action and figures (c) and
(d) represent the ghost-graviton interactions from the last two terms
in Eq. \eqref{seff1}.}\label{f3}
\end{figure}

The interaction vertices can also be derived directly from Eq. \eqref{seff1}.
Let us recall that the quantity ${S}_i$ represents all the
interaction terms, starting with the three-graviton vertex,
which arise from the expansion of the Einstein-Hilbert action in
powers of $\kappa$. Some of the expressions for the graviton
self-interaction vertices have been derived before up to the
five-graviton vertex \cite{Brandt:1992dk}.
Since the argument of ${S}_i$ in 
Eq. \eqref{seff1} has been shifted by a $\theta$ dependent quantity, 
there will also be additional interaction
terms between the $\theta$ field and the graviton.
In the figure \ref{f1} we show some of the new vertices involving the
this type of $\theta$-graviton interactions. The numbers inside the
blobs are meant to indicate when the
corresponding vertex comes from the cubic or quartic terms of ${S}_i$.
Here we are only considering the vertices which will contribute to the
one-loop graviton self-energy.
The third term in Eq. \eqref{seff1} also yields new $\theta$-graviton interactions.
In this case, there is only two diagrams which are shown in the figure \ref{f2}.
Finally, in the figure \ref{f3} we show the cubic and quartic graviton
self-interactions, as well as the two diagrams involving the
interaction of the graviton with the two types of Fermionic ghosts.
In the next section all the vertices shown in the above figures will
be employed in order to obtain the known gauge invariant result for the
leading high temperature limit of the static graviton self-energy.

\section{The static self-energy at finite temperature}
\begin{figure}[h!]
\begin{center}
\[
\begin{array}{cc}
\includegraphics[scale=0.4]{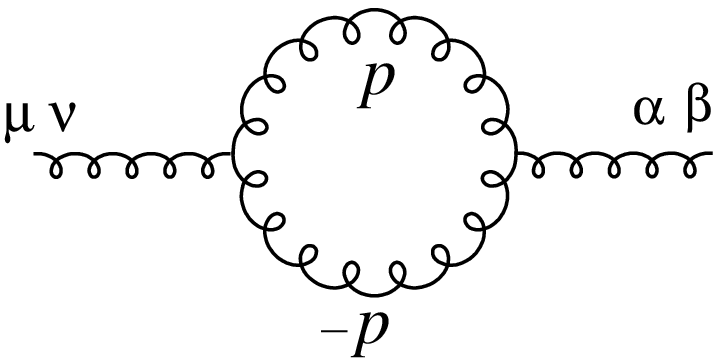} &
\qquad \includegraphics[scale=0.4]{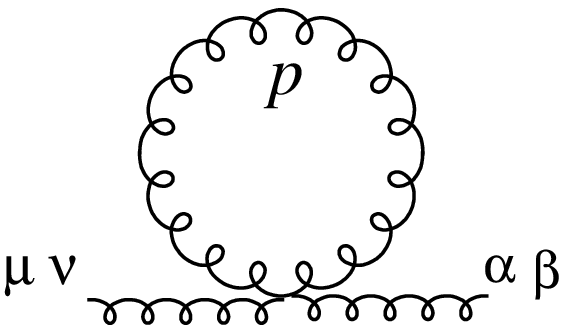}
\\ &  \\ (a) & \qquad (b) 
\\ & \\ & \\
\includegraphics[scale=0.4]{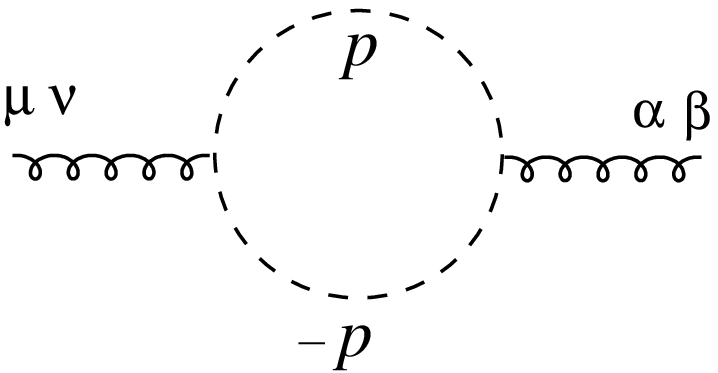}  &
\qquad \includegraphics[scale=0.4]{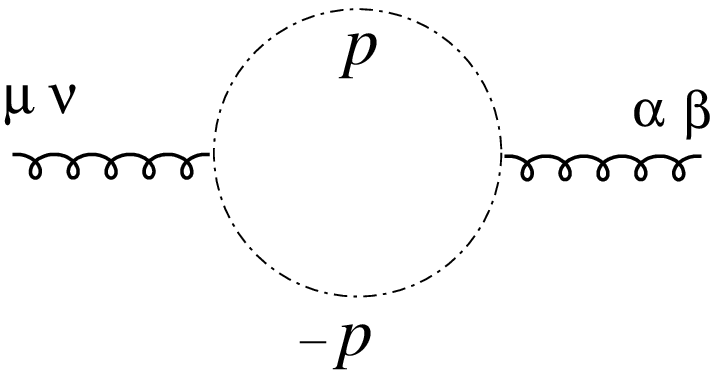}  
\\ &  \\ (c) & \qquad (d) 
\\ & \\ & \\
\includegraphics[scale=0.4]{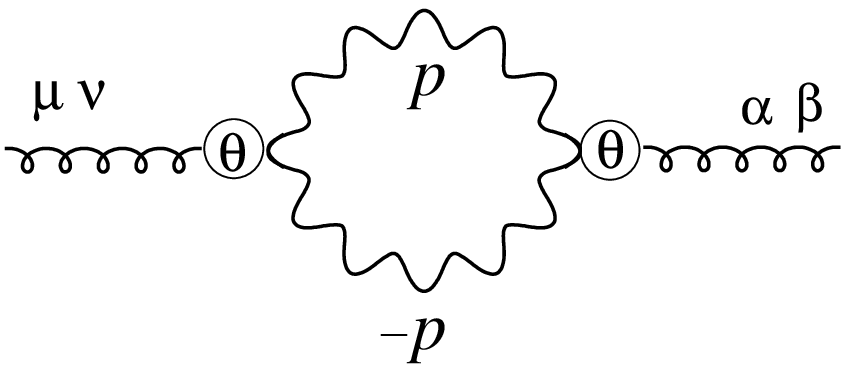}  &
\qquad \includegraphics[scale=0.4]{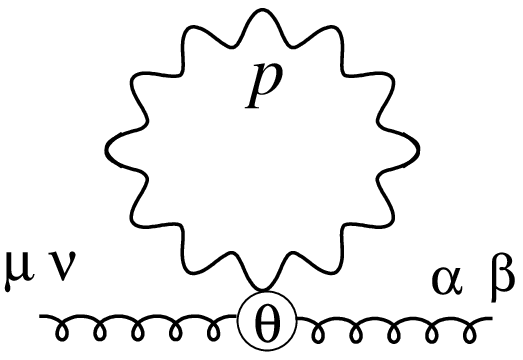}  
\\ &  \\ (e) & \qquad (f) 
\end{array}
\]
\end{center}
\caption{Diagrams which contribute the static limit of the graviton
self-energy. The curly and wave lines represent respectively
gravitons and the $\theta$ ghost. The dashed and
dot-dashed lines represent the two types Fermionic ghosts. Graphs (a),
(b), (e) and (f) have a symmetry factor $1/2$. A factor of $(-1)$ is
associated with the Fermionic ghost loops in figures (c) and (d).}\label{self_eh}
\end{figure}

\begin{figure}
\begin{center}
\[
\begin{array}{cc}
\includegraphics[scale=0.4]{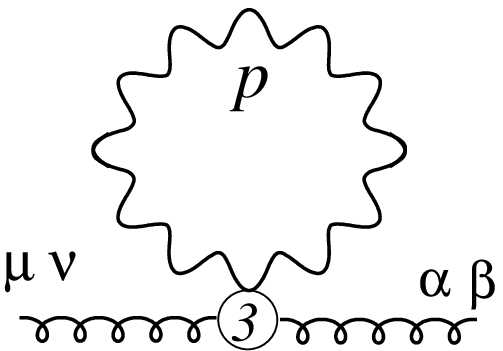}  &
\qquad \includegraphics[scale=0.4]{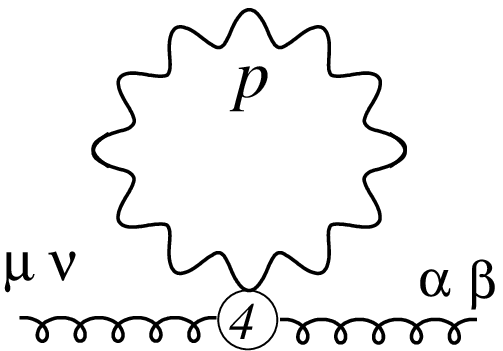}  
\\ &  \\ (g) & \qquad (h) 
\\ & \\ & \\
\includegraphics[scale=0.4]{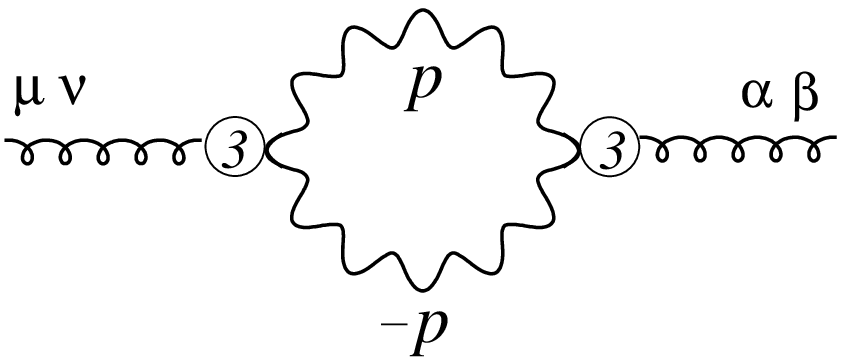}  &
\qquad \includegraphics[scale=0.4]{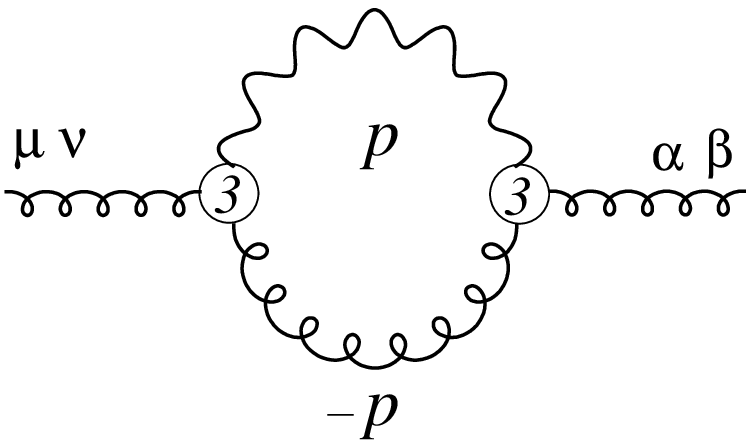}  
\\ &  \\ (i) & \qquad (j) 
\\ & \\ & \\
\includegraphics[scale=0.4]{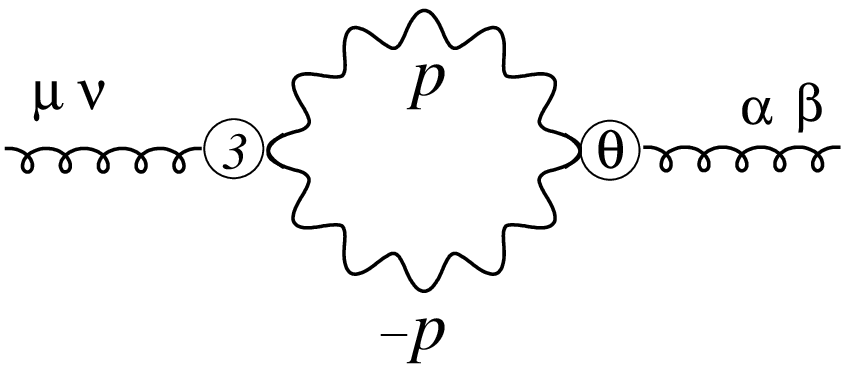}  &
\qquad \includegraphics[scale=0.4]{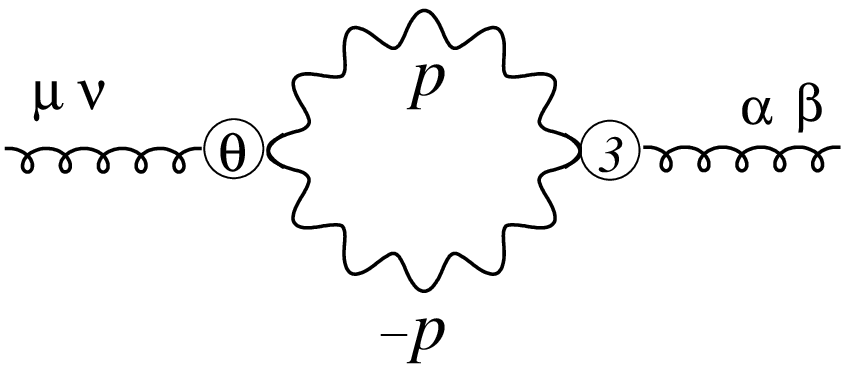}  
\\ &  \\ (k) & \qquad (l) 
\end{array}
%\nonumber
%\end{eqnarray}
\]
\end{center}
\caption{Contributions involving of the $\theta$ ghosts (wavy lines) 
to the static limit of the graviton self-energy.
Except for graph (j), all graphs have a symmetry factor $1/2$.
}\label{self_theta}
\end{figure}
The static graviton self-energy at finite temperature is a well known
quantity and one of the simplest examples exhibiting gauge independence
\cite{Rebhan:1990yr,Brandt:1998hd}. Therefore, it can
be used as a rather non-trivial test of the gauge fixing procedure
presented in the previous section.
% and, consequently, it will allow us
%to check the expressions for the vertices presented in the previous section.
In fact, there is an even simpler example, namely the
one-graviton function. This has been considered in
Ref. \cite{Brandt:2007td}, and used as a test of the vertices
involving only the interactions of three particles. In the case of the
self-energy, the contribution of all the vertices shown in the previous section will 
be taken into account as it can be seen in the figures 
\ref{self_eh} and \ref{self_theta}. 
Therefore, one of the main results of this section will be the verification of
the gauge independence of the static graviton self-energy, which will
be employed in the next section, together with the TT graviton
propagator, in order to derive the ressumation of the infrared
divergences of the free-energy.
%Except for the extra
%$d$ ghost, the graphs in figure \ref{self_eh} are the only ones that
%would arise in the deDonder gauge. On the other hand,
%all the graphs in figure \ref{self_theta} are only present in the
%non-quadratic gauge fixing procedure. 

Because of the algebraic complexity involved in the calculation of
some of these diagrams, we have considered the special case when the
external momentum vanishes. Nonetheless, for the dominant high
temperature contribution, this special choice has the
interesting property of being equivalent to the static limit, 
as we have found explicitly in the early stages of the present
investigation. More recently, this
equivalence of the static and the zero four-momenta limits
has been proved to be true for all the thermal Green's functions \cite{Frenkel:2009pi} 
(this is not so, however, for the long wavelenght limit \cite{Brandt:2009ht}). 
%This interesting property can also be verified
%directly comparing the zero external momentum result 
%with the static limit of the hard thermal loop self-energy
%obtained in Ref. \cite{Rebhan:1990yr}. 

All the zero momentum 
diagrams (a) to (j), as well as the sum of the diagrams (k) and (l),
in figures \ref{self_eh} and \ref{self_theta}
are such that their integrands have a tensor symmetry which allows one
to parametrize then in terms of the basis given
in Eqs. \eqref{base1}. Therefore, we can express 
the static thermal self-energy as
\begin{eqnarray}\label{self_gen1}
\Pi_{\mu\nu,\alpha\beta} &=& \sum_{I=1}^{11} \Pi^I_{\mu\nu,\alpha\beta}
\nonumber \\ &=&
\sum_{I=1}^{11} 
\int d^{d-1} p
\nonumber \\
&\times& T\!\!\!\! \sum_{n=-\infty}^{\infty}
\left(\sum_{i=1}^{5} C^i_I T^i_{\mu\nu\alpha\beta}(p)\right) ,
%\nonumber \\ & &
\end{eqnarray}
where $I=1,\cdots,10$ and $I=11$ labels the contributions of the diagrams from (a)
to (j)  and the one from the sum of diagrams (k) and (l), respectively.
We are using the imaginary time formalism
\cite{kapusta:book89,lebellac:book96,das:book97}, so that
the integrand depends on $n$ through the Matsubara frequencies $p_0=2\pi n T$.

Once we compute the contributions to integrands of $\Pi^I_{\mu\nu,\alpha\beta}$, 
the expressions for $C^i_I$ can be obtained in a straightforward manner
contracting the integrand of $\Pi^I_{\mu\nu,\alpha\beta}$ with the five tensors  
$T^i_{\mu\nu\alpha\beta}$ and solving the system of five equations.
Notice that the momentum independence of the tensors
$T^1_{\mu\nu\alpha\beta}$ and $T^2_{\mu\nu\alpha\beta}$ may require a prescription such as
\begin{eqnarray}\label{eq0}
\int d^{n-1} p \sum_{n=-\infty}^{\infty} 1 &=& \int d^{n-1} p \lim_{\epsilon\rightarrow 0}
\left(1+2 \sum_{n=0}^{\infty} \frac{1}{n^\epsilon}\right) 
\nonumber \\
&=&\int d^{n-1} p \left(1+2\zeta(0)\right) = 0.
\end{eqnarray}
However, for our present purpose, it would be interesting if the quantities 
\be
C^i = \sum_{I=1}^{11} C^i_I
\ee
happen to be gauge independent even before the sum and integration is performed.
In order to investigate this possibility, let us first write down the
results which one would obtain when the static self-energy is directly
computed in the deDonder gauge. The calculation is much simpler in
this case; only the diagrams (a) and (b) as well as one of the Fermionic 
ghost loops contribute to $C^i$. A straightforward calculation yields
%DeDonder := [cDD1 = 1/16 (dd - 5) dd, cDD2 = 0, cDD3 = - 1/8 dd (- 3 + dd),
%
%      cDD4 = 0, cDD5 = 1/4 dd (- 3 + dd)]
\begin{eqnarray}\label{dedonder1}
 C^1 &=& \frac{d(d-5)}{16} -\frac 1 2,\;
 C^2 =  C^4 = 0,\;
\nonumber \\
 C^3 &=& -\frac{d-3}{8},\;
 C^5 = -2  C^3 ,
\end{eqnarray}
where the second term in $C^1$ is the only non-vanishing contribution which
comes from the ghost loop.

Let us now consider the individual contributions of the diagrams in
figures \ref{self_eh} and \ref{self_theta} which arise in the case of
a general gauge fixing. The simplest diagrams 
are the ones shown in figures \ref{self_eh} (c) and (d).
Using the results of the previous section, we find that each of
these ghost loops yield an identical contribution  such that
\be\label{gh1}
 C^1_{3} = C^1_{4} = -\frac 1 2 ,\;\;\; 
 C^i_{3} = C^i_{4} = 0,\;\;\; i=2,\cdots,5 .
\ee
The $\theta$ ghost diagrams  (e) and (f) in figure \ref{self_eh} have propagators
and vertices which depend on the three gauge parameters as 
described in the previous section. Despite this,
the calculation yields a result such that all
gauge parameter dependence cancels out and we are left with 
\be\label{th1}
C^1_{5} + C^1_{6} = \frac 1 2,\;\;\; 
C^i_{5} + C^i_{6} = 0,\;\;\; i=2,\cdots,5 .
\ee
Notice that the integrand of one of the ghost loops in Eq. \eqref{gh1} cancels
with the corresponding $\theta$-loop in Eq. \eqref{th1} in such a way that
$C^1_{3}+C^1_{4}+C^1_{5}+C^1_{6} = -1/2$, which is the same as the
result that one would obtain in the deDonder gauge for the single
Fermionic ghost loop.

Let us now introduce the quantities
\be
\Delta^i = \sum_{I=1}^{6} C^i_I - C^i ,\;\;\; i=1,\cdots,5 ,
\ee
where $C^i$ are given by \eqref{dedonder1}. 
%This is simply the
%diference between the sum of the graphs in figure \ref{self_eh} and
%the gauge invariant result given 
If the gauge independence manifests at the integrand level,
then 
\be\label{gauge_inv1}
\Delta^i + \sum_{I=7}^{11} C^i_I = 0 ,\;\;\; i=1,\cdots,5
\ee 
should be verified. In the appendix we display the results for 
$\Delta^i$ and $C^i_I$ ($I=7,\cdots,11$) from these results it can be
verified that \eqref{gauge_inv1} is indeed satisfied. The expressions
in the appendix shows how individual diagrams can have a very involved
dependence on the three gauge parameters $\alpha$, $g_1$ and
$g_2$. This constitutes a rather non-trivial example of the
consistence of the general gauge fixing procedure presented in the
previous section.

Finally, let us substitute the gauge invariant results given in 
Eq. \eqref{dedonder1} into
Eq. \eqref{self_gen1} and perform the sum and integration. This 
straightforward and standard calculation yields the following temperature
dependent expression for the static graviton self-energy in $d$ space-time dimensions
\be\label{self_old}
\Pi^{term.}_{\mu\nu,\alpha\beta} = \sum_{i=1}^5 {\cal C}^iT^i_{\mu\nu,\alpha\beta}(u), 
\ee
where $u$ is the heat bath four velocity and 
%\begin{subequations}
\begin{eqnarray}\label{ccold}
{\cal C}^1&=& \frac{d (d-3) \,2^{(4-d)}\,\pi^{\frac{7-d}{2}} }{ (d-1) \Gamma\left(\frac{d-1}{2}\right) }  
G \rho_d ,
 \nonumber \\
{\cal C}^2&=& -   {\cal C}^1,\nonumber \\
%\frac{(d-3) d }{(d-1)
%\Gamma\left(\frac{d-1}{2}\right)2^{(d-4)} \pi^{\frac{d-7}{2}} }G \rho_d \\
{\cal C}^3&=& 0  \\
{\cal C}^4&=&  d\, {\cal C}^1, \nonumber \\
%\frac{(d-3)d^2}{ (d-1)
%  \Gamma\left(\frac{d-1}{2}\right) 2^{(d-4)} \pi^{\frac{d-7}{2}}} G\rho_d \\
{\cal C}^5&=& -d(d+2)\, {\cal C}^1 ; \;\;\;\;
\rho_d \equiv \frac{\Gamma(d) \zeta(d)}{\pi^2} T^d,\nonumber
%-  \frac{(d-3) d^2 (d+2) }{(d-1)
%  \Gamma\left(\frac{d-1}{2}\right) 2^{(d-4)} \pi^{\frac{d-7}{2}} } G \rho_d ;
\end{eqnarray}
%\end{subequations}
with $\Gamma$ and $\zeta$ respectively the the Euler's and Riemann's functions.
From these expressions we obtain, for $d=4$, the  results given
in Eqs. (3.8) of Ref. \cite{Rebhan:1990yr} (notice that in the present
work we have labeled the tensors $T^i$ in a different order, so that
the constants $c_i$ of Ref. \cite{Rebhan:1990yr} are such that
$c_1={\cal C}^1/\kappa^2 = \rho_4/12$,  $c_2={\cal C}^3=0$,  
$c_3={\cal C}^5/\kappa^2 =-2\rho_4$,
$c_4={\cal C}^2/\kappa^2 =-\rho_4/12$ and
$c_5={\cal C}^4/\kappa^2= \rho_4/3$  when $d=4$).  
It is also possible to include contributions of other thermal
particles, such as scalars or fermions. Such contributions will only
modify all the ${\cal C}^i$ by a common integer factor which counts
the number of degrees freedom associated with each field.

\section{Resummation of infrared divergences in the free-energy}
%The sum of all the infrared divergent contributions to the 
%free energy coming from the gravitons 
%in the TT gauge can be represented graphically as...
The sum of all the infrared divergent contributions to the
free-energy can be represented graphically as
\be\label{eq1}
\Omega(T) =
-\frac{1}{2}\left[
\frac{1}{2}\begin{array}{c}\includegraphics[scale=0.3]{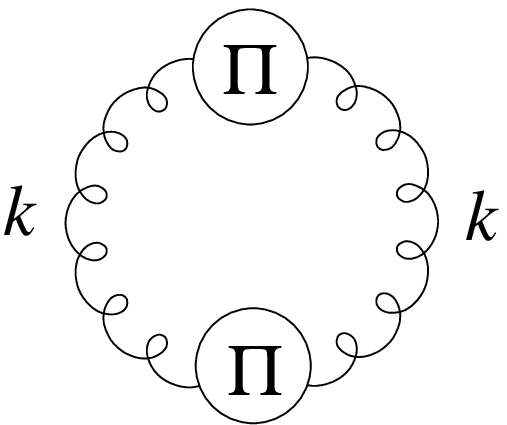}\end{array}+
\frac{1}{3}\begin{array}{c}\includegraphics[scale=0.3]{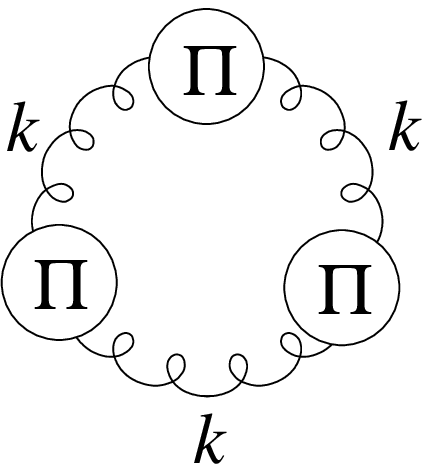}\end{array}
+\cdots 
\right],
\ee
where the quantity $\Pi$ stands for the static graviton self-energy
computed in the previous section. It is implicit in
\eqref{eq1} that we are considering 
only the {\it zero mode} contribution 
of the graviton propagator (denoted by curly lines),  
so that $k^2 = -|\vec k|^2$.  The reason for this is because each
photon propagator in \eqref{eq1} introduces a factor
\be
\frac{1}{k^2} = -\frac{1}{(2\pi n\, T)^2 + |\vec k|^2}, 
\ee
so that the zero mode ($n=0$)  of each individual ring diagram in \eqref{eq1}
yields an infrared divergent contribution, when $d=4$, from the
momentum integration $\int d^3 k\cdots$.
In scalar theories, as well as in gauge
theories of spin one fields, it is well known that the sum of all such
infrared divergent contributions yields a finite result which is
non-analytic in the coupling constant \cite{lebellac:book96}. 
In what follows we will employ the results of the last two sections in
order to investigate how a similar result may be achieved 
in the case of spin two fields. The details of this analysis show
explicitly that the use of the TT graviton propagator allows one to
obtain an explicit and compact result.
%We note here that the ghost self-energies are 
%sub-leading at high temperature, being of order $G T^2 k^2$.
%Due to extra factor of $k^2$, the static contributions to
%the ring diagrams coming from the ghost propagators will
%be convergent in the infrared region, and may therefore
%be neglected for our purpose.

Since we are going to employ the TT graviton propagator, the
computation of the right hand side of Eq. \eqref{eq1} becomes much
simpler if we express the static self-energy in terms of its traceless and
transverse components. However, at finite temperature there are three
distinct TT tensors tensors which may depend on the momentum $k$ and
the heat bath four velocity $u$. In the static case, 
when $k\cdot u=k_0=0$ (we are adopting the rest frame of the heat bath), 
these tensors can be written in $d$ dimensions as
\begin{subequations}\label{TTT}
\begin{eqnarray}
T_{\mu \nu \lambda \sigma}^A(u,k) &=&
\frac{1}{2}\left(\overline{\eta}_{\mu \lambda} \overline{\eta}_{\nu
    \sigma}+\overline{\eta}_{\nu \lambda}\overline{\eta}_{\mu \sigma}
\right. \nonumber \\ 
&& \qquad -
\left.\frac{2}{d-2}\overline{\eta}_{\mu \nu}\overline{\eta}_{\lambda \sigma}\right) \\
T_{\mu \nu \lambda \sigma}^B(u,k)&=& \frac{1}{2} P_{\mu \lambda}
P_{\nu \sigma}+ \frac{1}{2}P_{\nu \lambda} P_{\mu \sigma}
\nonumber \\ &-&\frac{1}{d-1}P_{\mu \nu} P_{\lambda \sigma} 
\nonumber \\
&-&T_{\mu \nu \lambda \sigma}^A(u,k) - T_{\mu \nu \lambda\sigma}^C(u,k) \\ 
T_{\mu \nu \lambda \sigma}^C(u,k) &=&  \frac{1}{(d-1)(d-2)}
\nonumber \\ &\times&
\left[(d-1) u_{\mu}
  u_{\nu}+ \frac{k_{\mu} k_{\nu}}{k^2} - \eta_{\mu \nu}\right]
\nonumber \\ &&
\left[(d-1) u_{\lambda} u_{\sigma}+ \frac{k_{\lambda} k_{\sigma}}{k^2} -
  \eta_{\lambda \sigma}\right] ,
\end{eqnarray}
\end{subequations}
where
\be
\overline{\eta}_{\mu \nu}=\eta_{\mu \nu}-u_{\mu}u_{\nu}-\frac{k_{\mu}k_{\nu}}{k^2}.
\ee
For $d=4$ the above expressions reduces to the static limit of the
corresponding ones given in Ref. \cite{Rebhan:1990yr}.
Notice that the sum of the three traceless transverse tensors
coincides with the numerator of the propagator in
Eq. \eqref{propTT}, which represents the only traceless transverse
tensor available at zero temperature, so that we can write
\be\label{DTT}
{D}^{TT}_{\mu \nu ; \alpha \beta} = \frac{1}{k^2}\left(
T_{\mu \nu \lambda \sigma}^A+T_{\mu \nu \lambda \sigma}^B+
T_{\mu \nu \lambda \sigma}^C\right),
\ee

In terms of its TT components the dominant contribution to the
static self-energy can then be expressed as
\be\label{selfTT}
\Pi^{term.}_{\mu\nu,\alpha\beta} = \sum_{I=A,B,C} \bar{\cal C}^I
T^I_{\mu\nu,\alpha\beta} + \bar{\cal C}^4 T^4_{\mu\nu,\alpha\beta}+ \cdots ,
%+ \bar{\cal C}^2T^2_{\mu\nu,\alpha\beta} + \bar{\cal C}^4 T^4_{\mu\nu,\alpha\beta},
\ee
where the ellipsis represents terms which are orthogonal to the TT
tensors and the constants $\bar{\cal C}^I$ and $\bar{\cal C}^4$
can be solved in terms the ones given in Eq. \eqref{ccold} as follows
\begin{subequations}\label{solctedtt}
\begin{eqnarray}
\bar{\cal C}^A&=&\bar{\cal C}^B=2{\cal C}^1\\
\bar{\cal C}^C&=&2{\cal C}^1 + \frac{d-2}{d-1}{\cal C}^5 \\
\bar{\cal C}^4&=&{\cal C}^4 + \frac{{\cal C}^5}{d-1}.
%\bar{\cal C}^4&=&\frac{2}{d}{\cal C}^1 + {\cal C}^2 -\frac{{\cal C}^5}{d^2} .
%\bar{\cal C}^4&=&\frac{{\cal C}^5}{d}+{\cal C}^4.
\end{eqnarray}
\end{subequations}

Notice that the dependence of each individual TT tensors $T^I$ on the
momentum variable $k=(0,\vec k)$  
is canceled when all the terms in \eqref{selfTT} are taken into account 
so that the result becomes
identical to the one in Eq. \eqref{self_old}. This apparent unnecessary
complication is very convenient in order to perform the
calculation of the ring diagrams in \eqref{eq1}.
This happens because the TT tensors in \eqref{TTT} are not only
traceless and transverse, but also enjoy of some very
important properties. First, they are idempotent so that
\be
{T}^I_{\mu\nu\;\lambda\delta} {{T}^I}^{\lambda\delta}_{\;\;\;\;\alpha\beta}
={T}^I_{\mu\nu\;\alpha\beta};\;\;\; (I=A, B, C). 
\ee
They are also orthogonal
\be
{T}^I_{\mu\nu\;\lambda\delta} {{T}^{I^\prime}}^{\lambda\delta}_{\;\;\;\;\alpha\beta}
=0;\;\;\; I\neq I^\prime
\ee
Their ``norm'' is given by
\begin{subequations}
\begin{eqnarray}
{T}^A_{\mu\nu\;\lambda\delta} {{T}^{A}}^{\mu\nu\;\lambda\delta}
&=&\frac{d(d-3)}{2} \\
{T}^B_{\mu\nu\;\lambda\delta} {{T}^{B}}^{\mu\nu\;\lambda\delta}
&=&d-2 \\
{T}^C_{\mu\nu\;\lambda\delta} {{T}^{C}}^{\mu\nu\;\lambda\delta}
&=& 1.
\end{eqnarray}
\end{subequations}
Finally, the tensors $T^A$ and $T^B$ also satisfies
\be
{T}^A_{\mu\nu\;\lambda\delta} u^\lambda=
{T}^B_{\mu\nu\;\lambda\delta} u^\lambda u^\delta =0.
\ee
Using these properties, as well as Eqs. \eqref{DTT} and 
\eqref{selfTT} it is straightforward to show that
the integrand of the first ring diagram in Eq. \eqref{eq1} is given by
\begin{eqnarray}
D^{TT\;\mu\nu \mu_1\nu_1} \Pi^{term.}_{\mu_1\nu_1 \lambda_1\sigma_1} 
D^{TT\;\lambda_1\sigma_1 \mu_2\nu_2}\Pi^{term.}_{\mu_2\nu_2 \mu\nu} 
\nonumber \\
=\frac{1}{k^4} \left(\frac{d(d-3)}{2} {(\bar{\cal C}^A)}^2
+(d-2){(\bar{\cal C}^B)}^2+{(\bar{\cal C}^C)}^2\right). 
\nonumber \\
\end{eqnarray}
Similarly, in the case of the higher order graphs we obtain
\begin{eqnarray}\label{genintfe}
D^{TT\;\mu\nu \mu_1\nu_1} \Pi^{term.}_{\mu_1\nu_1 \lambda_1\sigma_1} 
\cdots
D^{TT\;\lambda_{n-1}\sigma_{n-1} \mu_n\nu_n}\Pi^{term.}_{\mu_n\nu_n \mu\nu} 
\nonumber \\ 
=\frac{1}{(-|\vec k|^2)^n} \left(\frac{d(d-3)}{2} {(\bar{\cal C}^A)}^n
+(d-2){(\bar{\cal C}^B)}^n+{(\bar{\cal C}^C)}^n\right) ,
\nonumber \\ 
\end{eqnarray}
where we have used the zero mode condition $k^2=-|\vec k|^2$. 
Notice that the components of the self-energy which are not traceless
and transverse drops out in the final result. This is consistent with
the fact that the TT components are physical.

We have now all the ingredients to compute the sum of the infrared divergent
contributions to the free-energy in Eq. \eqref{eq1}. From Eq. \eqref{genintfe}  
we can see that the sum of the ring diagrams is composed of three
similar structures, each one having the same form as
$-\sum_{n=2}^{\infty} {(-x)^n}/{n} = \log(1+x)-x$, so that
the free-energy can be written as
%\begin{widetext}
\begin{eqnarray}\label{1tudo5a}
\Omega(T) &=& \frac{T}{2} 
\frac{1}{(2\pi)^{d-1}}
\left[\frac{d(d-3)}{2} I(\bar{\cal C}^A)\right.
\nonumber \\
&+&\left.
(d-2) I(\bar{\cal C}^B)+I(\bar{\cal C}^C)\right],
\end{eqnarray}
%\end{widetext}
where
\begin{eqnarray}
I(c) \equiv \int{d}^{d-1} k 
\left[\log\left(1+\frac{c}{|\vec k|^2}\right)-\frac{c}{|\vec k|^2} \right]
\end{eqnarray}
is a familiar integral which arises also in the context of scalar
or vector fields and it can be done in a closed form.
Performing the $d-2$ angular integral and using  
the change of variable $z=|\vec k|/\sqrt{c}$, we obtain
\be
I(c) =
\frac{2(\pi c)^{\frac{d-1}{2}}}{\Gamma\left(\frac{d-1}{2}\right)}
%c^{\frac{d-1}{2}} 
\int_0^\infty dz z^{d-2}\left[\log\left(1+\frac{1}{z^2}\right)
-\frac{1}{z^2}\right].
\ee
Using integration by parts, and performing the resulting integral yields
\begin{eqnarray}\label{xxx2}
I(c) &=& -\frac{4(\pi c)^{\frac{d-1}{2}}}{(d-1)\Gamma\left(\frac{d-1}{2}\right)}
%-\frac{2 ?(d) c^{\frac{d-1}{2}}}{d-1} 
\int_0^\infty dz \frac{z^{d-4}}{z^2+1} \nonumber \\
&=& 
-\frac{4\Gamma\left(\frac{5-d}{2}\right)}{(d-1)(d-3)}(\pi c)^{\frac{d-1}{2}}.
%-\frac{\pi ?(d) }{d-1}\sec\left({\frac{(d-4)\pi}{2}}\right)c^{\frac{d-1}{2}} .
\end{eqnarray}
Substituting \eqref{xxx2} into \eqref{1tudo5a} we obtain the
following result
\begin{eqnarray}\label{1tudo5b}
\Omega(T) &=& -
\frac{2\Gamma\left(\frac{5-d}{2}\right) T}{(d-1)(d-3)(2\pi)^{d-1}}
\left[\frac{d(d-3)}{2} (\pi \bar{\cal C}^A)^{\frac{d-1}{2}}\right.
\nonumber \\
&+&\left.
(d-2) (\pi\bar{\cal C}^B)^{\frac{d-1}{2}}+(\pi\bar{\cal C}^C)^{\frac{d-1}{2}}\right].
\end{eqnarray}
Finally, using the Eqs. \eqref{ccold} and \eqref{solctedtt} this expression
can be written as
\begin{eqnarray}\label{1tudo5c}
\Omega(T) &=& -
\frac{2\Gamma\left(\frac{5-d}{2}\right)} 
{(d-1)(d-3)(2\pi)^{d-1}}  
\nonumber \\  &\times &
\left[
\frac{2^{(5-d)}\,\pi^{\frac{5-d}{2}}\Gamma(d+1)\zeta(d) }{\Gamma\left(\frac{d-1}{2}\right) }  
\right]^{\frac{d-1}{2}}
\nonumber \\  &\times &
\left\{\frac{d(d-3)}{2}\left(\frac{d-3}{d-1}\right)^{\frac{d-1}{2}} 
\right. \nonumber \\ & & \left.
+ (d-2) \left(\frac{d-3}{d-1}\right)^{\frac{d-1}{2}} 
\right.\nonumber \\
&+&\left.
\left[\left(1-\frac{(d-2) d (d+2)}{2(d-1)} \right)\left(\frac{d-3}{d-1}\right)\right]^{\frac{d-1}{2}}
\right\} \nonumber \\
&\times & 
{\left(G T^{d-2}\right)^{\frac{d-1}{2}}T^d}  .
\end{eqnarray}

\section{Discussion}

In this work we have investigated the possibility of extending the
general gauge fixing procedure of Ref. \cite{Brandt:2007td} in order
to take into account interacting spin two fields. As an explicit
example, we have computed the
dominant one-loop contributions to the thermal self-energy of the
graviton, and verified that it agrees with the known gauge
invariant result. Then, using a decomposition of the
self-energy in terms of the three traceless and transverse tensors
which arises at finite temperature, as well as the TT graviton
propagator, we were able to obtain a closed form expression for the
sum of the infrared divergent contributions to the free energy. 

We note that the above gauge invariant result for the graviton
thermal self-energy satisfies a simple Ward identity 
\cite{Frenkel:1991dw} (not just
a BRST identity) which is a consequence of the fact that the ghost
self-energies are sub-leading at high temperature. Consequently, on
dimensional grounds, the thermal self-energies associated with the
$c_\mu$ and $d_\mu$ ghost fields must be of order 
$G T^{d-2} k^2$, whereas the $\theta_\mu$ 
ghost self-energy would be proportional to $G T^{d-2} k^4$, where the
factor $G T^{d-2}$ is dimensionless. The presence
of these powers of $k$  in the thermal ghost self-energies ensures
that infrared divergences will be absent in ring diagrams involving
ghost propagators. This justifies the neglect of such diagrams in
the evaluation of the leading infrared divergent contributions to
the free energy of spin-two fields.
(A similar behavior occurs also in the case of the free-energy in QCD
\cite{kapusta:book89}.)

The result presented in Eq. \eqref{1tudo5c} has some interesting
features which we would like to stress. First, for odd space-time
dimensions it is a real and singular function. On the other hand,
for even space-time dimensions, 
it is a finite and non-analytic function of %the dimensionless quantity 
$G T^{d-2}$ as one would expect for a non-perturbative quantity. 
However, in this case it acquires an imaginary part. 
For instance, for $d=4$ the third 
term inside the curly brackets, which can be traced back to the
$T^{C}$ component of the self-energy,  
is equal to $(-7/3)^{3/2}$. As a result, one would conclude that
the gravitational $C$-mode is unstable, since the imaginary part of
the free energy is connected with the decay rate of the quantum vacuum 
\cite{Affleck:1980ac}. However, a detailed investigation shows
that the graviton self-energy, which is proportional to $G T^4$, 
is of the same order as the solution of the Einstein 
equation for the curvature tensor, when the thermal energy momentum
tensor is taken into account. Therefore, by consistency, 
one should also take into account the curvature
corrections in the analysis of instabilities of gravity at finite
temperature. These corrections \cite{Rebhan:1990yr,Brandt:1998hd} have the effect of adding
some extra contributions to the self-energy in
such a way that the $C$-mode contribution to $\Omega(T)$ would 
change the third term of the curly bracket of Eq. \eqref{1tudo5c} to
$(-7/3+5/27)^{3/2}$, which is still imaginary. 
This term may be related to an imaginary value of a thermal Jeans mass
\cite{Rebhan:1990yr,Brandt:1998hd}, which reflects the instability of the system due
to the universal attractive nature of gravity.

\acknowledgements

F. T. Brandt and J. Frenkel would like to thank Fapesp and Cnpq for
financial support. 
J. B. Siqueira would like to thank Capes and Cnpq for financial support.
F. T. Brandt would like to thank A. Bessa for many helpful discussions.
{\hbox{D. G. C. McKeon}} would like to thank Roger MacLeod for a helpful suggestion.

\appendix

\section*{Appendix}
Here we display the expressions for the quantities introduced in 
Eq. \eqref{gauge_inv1}.
\begin{widetext}
\begin{subequations}
\begin{equation}
\Delta^1={\frac{\left (3\,d-7\right )\left ({g_1}-{g_2}\right )^{2}}{8\,\left[(d-1)(g_1-g_2)^2 + 2 \alpha (d-2)(g_1+1)(g_2+1)\right]}}
\end{equation}

\begin{equation}
C^1_{{7}}=0
\end{equation}

\begin{equation}
G^1_{{8}}=\frac{\left({d}^{2}-11\,d+22\right)\left({g_1}-{g_2}\right)^{2}}
{8\,\left(d-2\right)\,\left[(d-1)(g_1-g_2)^2 + 2 \alpha (d-2)(g_1+1)(g_2+1)\right]}
\end{equation}

\begin{equation}
G^1_{{9}}=0
\end{equation}

\begin{equation}
G^1_{{10}}=-{\frac{\left(d-3\right)^{2}\,\left ({g_1}-{g_2}\right )^{2}}{2\,\left (d-2\right)\,\left[(d-1)(g_1-g_2)^2 + 2 \alpha (d-2)(g_1+1)(g_2+1)\right]}}
\end{equation}

\begin{equation}
G^1_{{11}}=0
\end{equation}
\end{subequations}

\begin{subequations}
\begin{equation}
\Delta^2=-{\frac {\left ({g_1}-{g_2}\right )^{2}\,{
\left[4\,\left(d-3\right)\left({g_1}-{g_2}\right)^{2}+{\alpha}\,\left(d-1\right)\left(d-2\right)\left({g_1}+1\right)\left({g_2}+1\right)\right]
}}{4\,{\left[(d-1)(g_1-g_2)^2 + 2 \alpha (d-2)(g_1+1)(g_2+1)\right]}^{2}}}
\end{equation}

\begin{equation}
C^2_{{7}}=0
\end{equation}

\begin{equation}
C^2_{{8}}=-{\frac {\left (d^2-11\,d+26\right)\left({g_1}-{g_2}\right)^{2}}{8\,\left(d-2\right)\left[(d-1)(g_1-g_2)^2 + 2 \alpha (d-2)(g_1+1)(g_2+1)\right]}}
\end{equation}

\begin{equation}
C^2_{{9}}={\frac{\left(d-5\right)^{2}\left({g_1}-{g_2}\right )^{4}}{8\,{\left[(d-1)(g_1-g_2)^2 + 2 \alpha (d-2)(g_1+1)(g_2+1)\right]}^{2}}}
\end{equation}

\begin{equation}
C^2_{{10}}={\frac {\left ({g_1}-{g_2}\right )^{2}\,{
\left[2\,\left(d-3\right)^{2}\,\left({g_1}-{g_2}\right)^{2}+\alpha
\left(d-2\right )\left ({{d}}^{2}-7\,{d}+14\right)\left ({g_1}+1\right
)\left({g_2}+1\right)\right]
}}{2\,\left(d-2\right)\,{\left[(d-1)(g_1-g_2)^2 + 2 \alpha (d-2)(g_1+1)(g_2+1)\right]}^{2}}}
\end{equation}

\begin{equation}
C^2_{{11}}=0
\end{equation}
\end{subequations}

\begin{subequations}
\begin{equation}
\Delta^3=-{\frac {\left (d-3\right )\left ({g_1}-{g_2}\right )^{2}}{4\,\left[(d-1)(g_1-g_2)^2 + 2 \alpha (d-2)(g_1+1)(g_2+1)\right]}}
\end{equation}

\begin{equation}
C^3_{{7}}=0
\end{equation}

\begin{equation}
C^3_{{8}}=-\frac {\left (d-3\right )\left(d-4\right )\left({g_1}-{g_2}\right)^{2}}
{4\,\left(d-2\right)\,\left[(d-1)(g_1-g_2)^2 + 2 \alpha (d-2)(g_1+1)(g_2+1)\right]}
\end{equation}

\begin{equation}
C^3_{{9}}=0
\end{equation}

\begin{equation}
C^3_{{10}}={\frac{\left(d-3\right)^{2}\,\left({g_1}-{g_2}\right)^{2}}{2\,\left (d-2\right)\,\left[(d-1)(g_1-g_2)^2 + 2 \alpha (d-2)(g_1+1)(g_2+1)\right]}}
\end{equation}

\begin{equation}
C^3_{{11}}=0
\end{equation}
\end{subequations}

\begin{subequations}
\begin{equation}
\Delta^4=\frac {\left (d-3\right )\left ({g_1}-{g_2}\right )^{2}\,{
\left[2\,\left ({g_1}-{g_2}\right)^{2}+\alpha\left(d-2\right)\left({g_1}+1\right)\left ({g_2}+1\right)\right]
}}{2\,{\left[(d-1)(g_1-g_2)^2 + 2 \alpha (d-2)(g_1+1)(g_2+1)\right]}^{2}}
\end{equation}

\begin{equation}
C^4_{{7}}=-{\frac{\left(d-5\right)\left ({g_1}-{g_2}\right )^{2}}{2\,\left[(d-1)(g_1-g_2)^2 + 2 \alpha (d-2)(g_1+1)(g_2+1)\right]}}
\end{equation}

\begin{equation}
C^4_{{8}}=\frac{\left(d-6\right)\left(d-3\right)\left({g_1}-{g_2}\right)^{2}}{4\,\left(d-2\right)\,\left[(d-1)(g_1-g_2)^2 + 2 \alpha (d-2)(g_1+1)(g_2+1)\right]}
\end{equation}

\begin{equation}
C^4_{{9}}=-{\frac{\left(d-3\right)\left(d-5\right)\left({g_1}-{g_2}\right)^{4}}{4\,{\left[(d-1)(g_1-g_2)^2 + 2 \alpha (d-2)(g_1+1)(g_2+1)\right]}^{2}}}
\end{equation}

\begin{equation}
C^4_{{10}}=-{\frac {\left (d-3\right )\left
      ({g_1}-{g_2}\right)^{2}{
\left[\left (d-3\right )\left ({g_1}-{g_2}\right )^{2}+
\alpha\left(d-2\right )\left ({d}-4\right )\left({g_1}+1\right )\left
  ({g_2}+1\right )\right]
}}{\left (d-2\right){\left[(d-1)(g_1-g_2)^2 + 2 \alpha (d-2)(g_1+1)(g_2+1)\right]}^{2}}}
\end{equation}

\begin{equation}
C^4_{{11}}={\frac {\left(d-5\right)\left({g_1}-{g_2}\right )^{2}}{2\,\left[(d-1)(g_1-g_2)^2 + 2 \alpha (d-2)(g_1+1)(g_2+1)\right]}}
\end{equation}
\end{subequations}

\begin{subequations}
\begin{equation}
\Delta^5={\frac {\left (d-3\right )^{2}\left ({g_1}-{g_2}\right )^{4}}{2\,{\left[(d-1)(g_1-g_2)^2 + 2 \alpha (d-2)(g_1+1)(g_2+1)\right]}^{2}}}
\end{equation}

\begin{equation}
C^5_{{7}}={\frac {\left (2\,d-6\right )\left ({g_1}-{g_2}\right )^{2}}{\left[(d-1)(g_1-g_2)^2 + 2 \alpha (d-2)(g_1+1)(g_2+1)\right]}}
\end{equation}

\begin{equation}
C^5_{{8}}=0
\end{equation}

\begin{equation}
C^5_{{9}}={\frac{\left(d-3\right )^{2}\left({g_1}-{g_2}\right)^{4}}{2\,{\left[(d-1)(g_1-g_2)^2 + 2 \alpha (d-2)(g_1+1)(g_2+1)\right]}^{2}}}
\end{equation}

\begin{equation}
C^5_{{10}}=-{\frac{\left (d-3\right )^{2}\left ({g_1}-{g_2}\right)^{4}}{{\left[(d-1)(g_1-g_2)^2 + 2 \alpha (d-2)(g_1+1)(g_2+1)\right]}^{2}}}
\end{equation}

\begin{equation}
C^5_{{11}}=-{\frac {\left(2\,d-6\right )\left({g_1}-{g_2}\right )^{2}}{\left[(d-1)(g_1-g_2)^2 + 2 \alpha (d-2)(g_1+1)(g_2+1)\right]}}.
\end{equation}
\end{subequations}

\newpage
\end{widetext}

\end{document}